# Développement des méthodes de corrélation d'images pour l'analyse de la mobilité moléculaire dans un système biologique complexe spatialement et temporellement.

Chen Chen[1], Perrine Paul-Gilloteaux[1], Timothée Vignaud[2], Jean Salamero[1], François Waharte[1]

[1]PICT-IBiSA, UMR144 CNRS-Insitut Curie, 26 rue d'Ulm, 75248 Paris cedex 05
[2]Laboratoire de Physiologie Cellulaire et Végétale, Institut de Recherche en Technologies et Sciences pour le Vivant, CNRS/UJF/INRA/CEA, 17 rue des martyrs, 38054, Grenoble, France

*Jean.Salamero@curie.fr*

**Abstract**  *Development of methods for image correlation analysis of molecular mobility in a spatially and temporally complex biological system.*

Fluorescence spectroscopy is an image correlation technique to analyze and characterize the molecular dynamics from a sequence of fluorescence images. Many image correlation techniques have been developed for different applications [1]. But in practice the use of these techniques is often limited to a manually selected region of analysis where it is assumed that the observed molecules have homogeneous and constant behavior over time. Due to the spatial and temporal complexity of biological objects, this assumption is frequently at fault. It is then necessary to propose a robust method for discriminating the different behaviors over time from experience, as well as identification of the type of dynamics (diffusion or flow). This paper presents an original system of automatic discrimination and identification of spatially and temporally uniform regions for the analysis of molecular different dynamics over time by calculating STICS (Spatio-Temporal Image Correlation Spectroscopy) at different time lags. An evaluation of system performance is presented on simulated images and images acquired by fluorescence microscopy on actin cytoskeleton.

## 1    Introduction :

La compréhension de nombreux processus biologiques se produisant au sein des cellules nécessite d'être en mesure de déterminer la mobilité des protéines en jeu (diffusion, flux...), ceci dans les trois dimensions spatiales et à différentes échelles spatiales et temporelles. La microscopie de fluorescence est un outil reconnu pour l'analyse de la répartition spatiale de molécules marquées spécifiquement, à l'échelle sub-micrométrique. D'un autre côté, l'analyse des fluctuations de fluorescence au cours du temps dans un volume donné permet de mesurer la mobilité des molécules fluorescentes présentes dans l'échantillon. La combinaison de cette analyse avec l'imagerie microscopique de fluorescence a donné naissance à plusieurs techniques basées sur l'analyse des images pour la mesure de mobilité, mise en œuvre pour différentes applications [1]. Par exemple, la corrélation spatiale des images, technique ICS (en anglais: Image Correlation Spectroscopy) permet d'obtenir des informations sur la densité de molécules fluorescentes. La corrélation temporelle via le TICS (en anglais: Temporal ICS) permet d'extraire le coefficient de diffusion et la vitesse de flux. Enfin, la combinaison des corrélations spatiales temporelles STICS (en anglais : Spatial-Temporal ICS) donne également une mesure de la direction et la vitesse de déplacement des molécules fluorescentes. En raison de la complexité des systèmes biologiques, les molécules n'ont pas forcément un comportement dynamique homogène spatialement, ni constant au cours du temps. La région spatio-temporelle pour l'analyse de la dynamique est actuellement choisie manuellement dans une séquence d'images fluorescences. Dans cet article, nous proposons un système original qui est basé sur le calcul de STICS à différents décalages temporels, afin d'identifier automatiquement le comportement dynamique spatio-temporelle dans la séquence d'images de fluorescence (flux ou diffusion).

Dans la suite de cet article, nous introduisons en section 2 la théorie de spectroscopie de corrélation d'image spatio-temporelle (en anglais: STICS). Notre contribution, consistant en un système d'identification automatique des régions uniformes, est présentée en section 3. La section 4 présente l'évaluation de notre système sur données simulées puis son application sur données expérimentales. Les images simulées ont l'avantage de permettre une évaluation quantitative de la méthode alors que les images expérimentales permettent de la mettre à l'épreuve sur des signaux traduisant des processus biologiques complexes (flux, diffusion, assemblage, désassemblage, agrégation, …).

La section 5 donne la conclusion et les perspectives de ce travail.

## 2. Méthode de spectroscopie de corrélation d'image spatio-temporelle (STICS) :

La méthode de STICS consiste à calculer la fonction de corrélation spatio-temporelle en fonction du temps. L'analyse spatio-temporelle permet d'observer la direction et la vitesse de déplacement de molécules fluorescentes. La fonction d'autocorrélation est définie par :

$$r(\xi, \eta, \tau) = \frac{<\Delta I(x,y,t) * \Delta I(x+\xi, y+\xi, t+\tau)>}{<I>_t * <I>_{t+\tau}} \quad (1)$$

Où : ξ et η représentent le décalage spatial, τ le décalage temporel. $\Delta I$ est la variation d'intensité de pixel $(x,y)$ au temps t, définie par $\Delta I(x,y,t) = I(x,y,t) - <I(x,y,t)>_t$. $<\bullet>$ et $<\bullet>_t$ correspondent respectivement à l'intensité moyenne d'image en t et t+τ.

Nous calculons la fonction de corrélation spatiale à différents décalages par Transformée de Fourier (figure 1). Un filtre moyenneur permet de filtrer les molécules immobiles. Enfin, nous estimons le déplacement du pic de $r$ par ajustement d'une fonction gaussienne 2D:

$$r(\xi, \eta, \tau) = r(\xi_p(\tau), \eta_p(\tau), \tau) \exp\left[-\frac{(\xi_p(\tau))^2 + (\eta_p(\tau))^2}{\omega_0^2}\right] + r_\infty(\tau) \quad (2)$$

Où : $\xi_p(\tau) et \eta_p(\tau)$ est la position de la valeur maximum du pic au décalage $\tau$. $r_\infty(\tau)$ est une compensation du niveau de base spatial. $\omega_0$ correspond à la taille du faisceau de la lumière d'excitation. Le changement des coordonnées du pic au cours du temps (s'il y a un flux) est ajusté par régression linéaire.

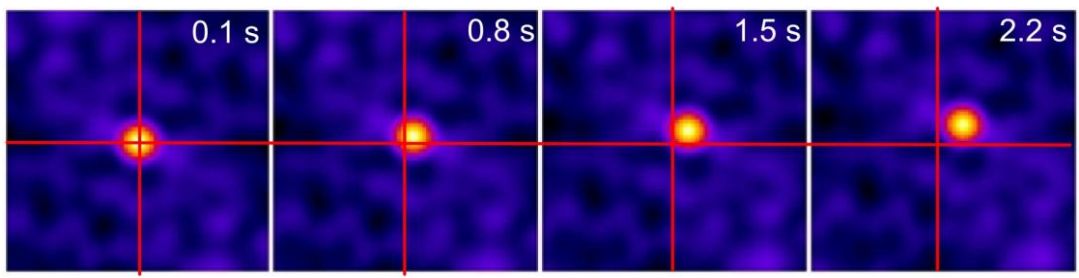

*Figure 1 : Allure de la fonction de corrélation spatiale à différents décalages temporels sans la population immobile.*

## 3. Système d'analyse par corrélation d'images :

En raison de la variation spatiale et temporelle des mécanismes moléculaires à l'intérieur d'une cellule biologique, ces techniques ne sont utilisables que pour analyser un comportement dynamique homogène. Elles ne sont pas forcément fiables pour une séquence d'image non homogène spatialement et/ou temporellement. Actuellement, les analyses sont souvent faites avec un choix arbitraire de la région spatio-temporelle pour mesurer la dynamique et des paramètres des modèles. Pour dépasser cette limitation forte, nous proposons dans cette section un système complet en appliquant la technique de STICS pour identifier automatiquement le nombre de comportement au cours du temps et mesurer la dynamique a posteriori, dont le schéma fonctionnel est représenté figure 2.

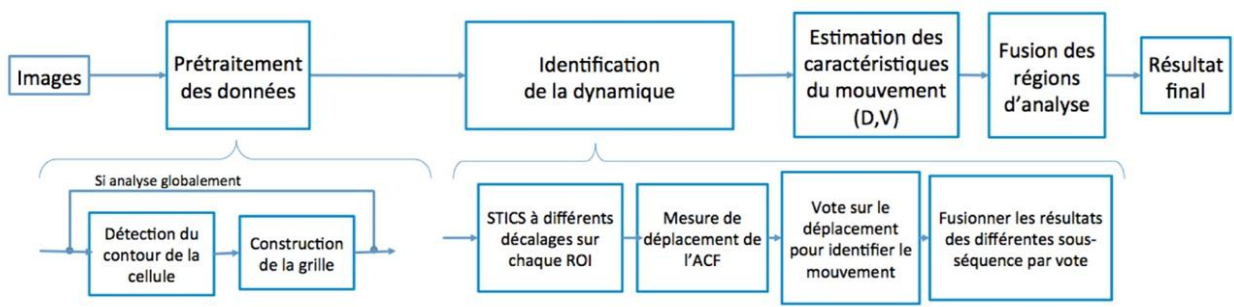

*Figure 2 : Schéma récapitulatif de la méthode proposée.*

Ce système comporte différentes étapes détaillées dans les sections suivantes mais l'ensemble est automatisé afin de ne pas requérir d'intervention de l'opérateur.

### *3.1 Prétraitements:*

La localisation et le choix de la région d'analyse est une étape importante pour mesurer les informations spatio-temporelles. Nous proposons une méthode de construction de grille en découpant l'image en sous-régions de même taille. La grille est basée sur une construction locale et globale. La détection des contours de la cellule par seuillage automatique permet de la localiser et de construire une grille à partir de l'objet détecté pour optimiser le temps de calcul. La grille peut aussi être construite sur l'ensemble de l'image lorsqu'on a besoin d'analyser les informations globalement. La taille initiale des régions doit être un compromis entre la précision [2] et le temps de calcul.

Nous montrons deux exemples dans la Figure 3, avec une grille globale pour les images simulées et locale pour un exemple de données expérimentales, qui sera commentée dans la section 4.

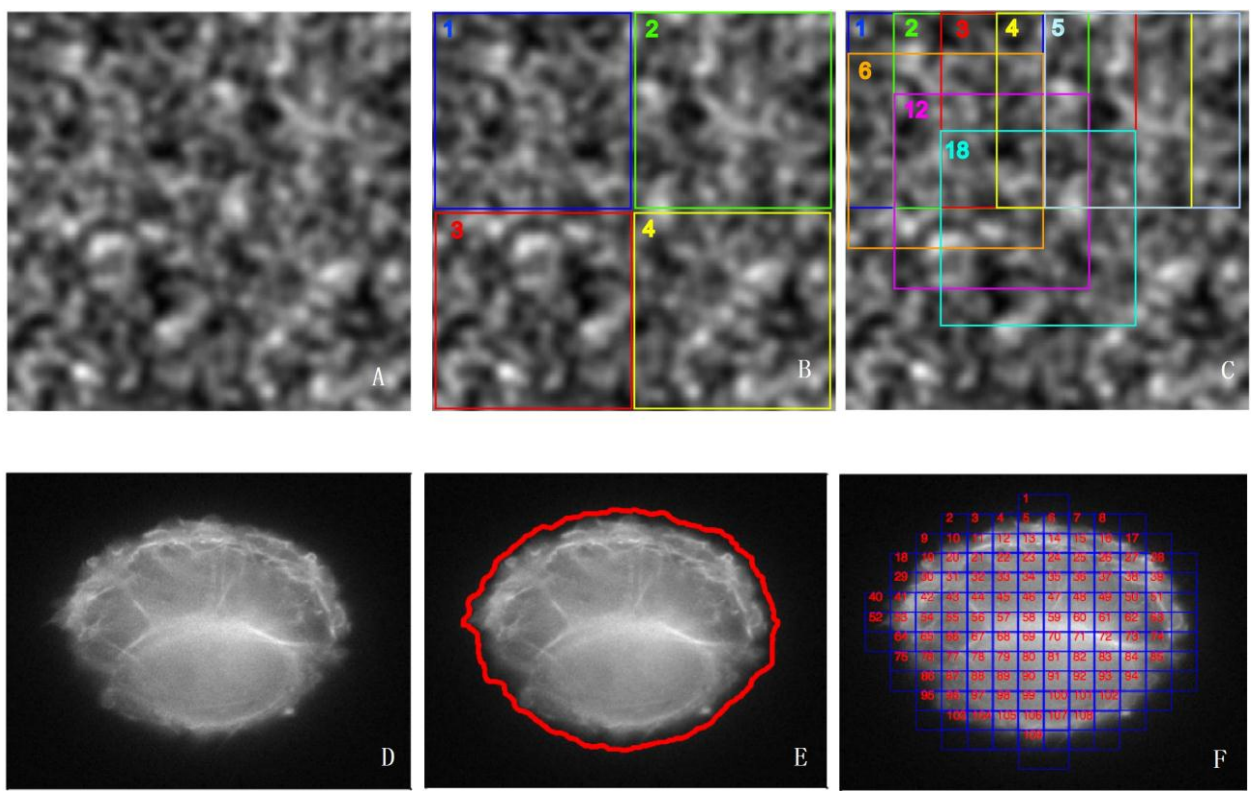

*Figure 3 : Construction de la grille locale et globale. (A) image simulée (128*128 pixels). (B) Construction de la grille globale sans décalage et (C) avec décalage (D) Cellule marquée lifeact sur motif adhésif, permettant de se concentrer sur les mouvements sub-cellulaires sans mouvement ou déformation de la cellule (voir section 4). (E) Détection de contour. (F) Construction de la grille locale. La taille de sous-région est fixée à 64*64 pixels pour les 2 exemples, le décalage est de 16 pixels et de 32 pixels pour chaque exemple respectivement.*

### *3.2 Identification du type de dynamique par sous-séquence:*

Le choix de caractéristique doit permettre d'identifier des dynamiques différentes. Le processus de transport moléculaire peut induire deux types de dynamique différents. La diffusion moléculaire correspond à un mouvement passif, lié à l'agitation thermique des molécules du solvant. Au contraire, le transport par moteurs moléculaires induit un flux qui représente un mouvement dirigé actif. Le principe de l'identification de la dynamique que nous proposons est basé sur l'étude des profils d'amplitude en x et y de la corrélation $r$ de l'équation (2) sur de courtes séquences temporelles, comme montré sur la figure 4. Les valeurs de décalages spatiaux qui maximisent la corrélation sont alors comparés pour plusieurs décalages temporels et sont attribués à trois catégories : différence de décalages spatiaux positive, négative ou nulle (en pratique inférieur à un pixel). Une direction approximative est attribuée à chaque sous-séquence en considérant uniquement les catégories ayant reçu le plus de vote, pour x et pour y. Une différence nulle en x et y est attribuée à une dynamique de type diffusion, le reste est considéré comme du flux, ce qui permet de choisir le modèle le plus adapté. Une catégorie supplémentaire de dynamique inconnue reçoit les sous-séquences qui n'ont pas atteint de majorité franche.

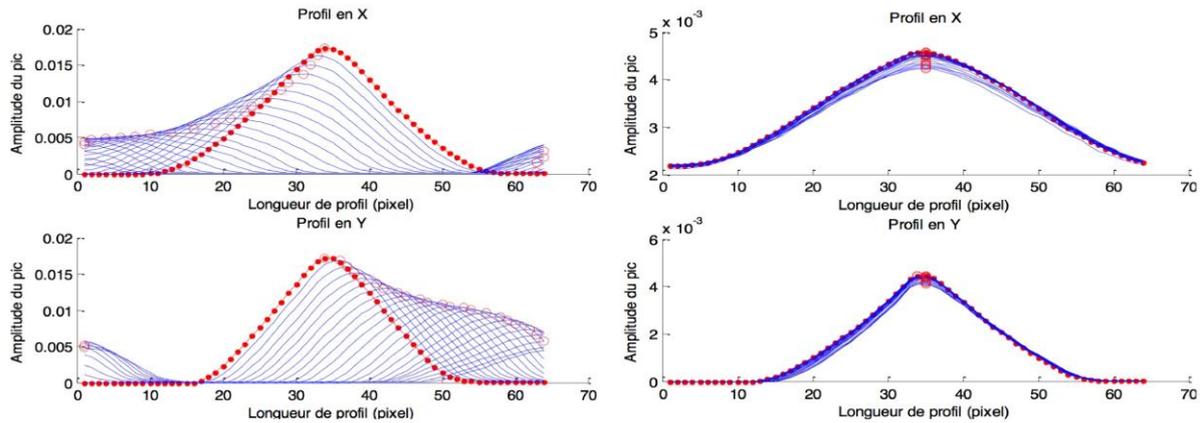

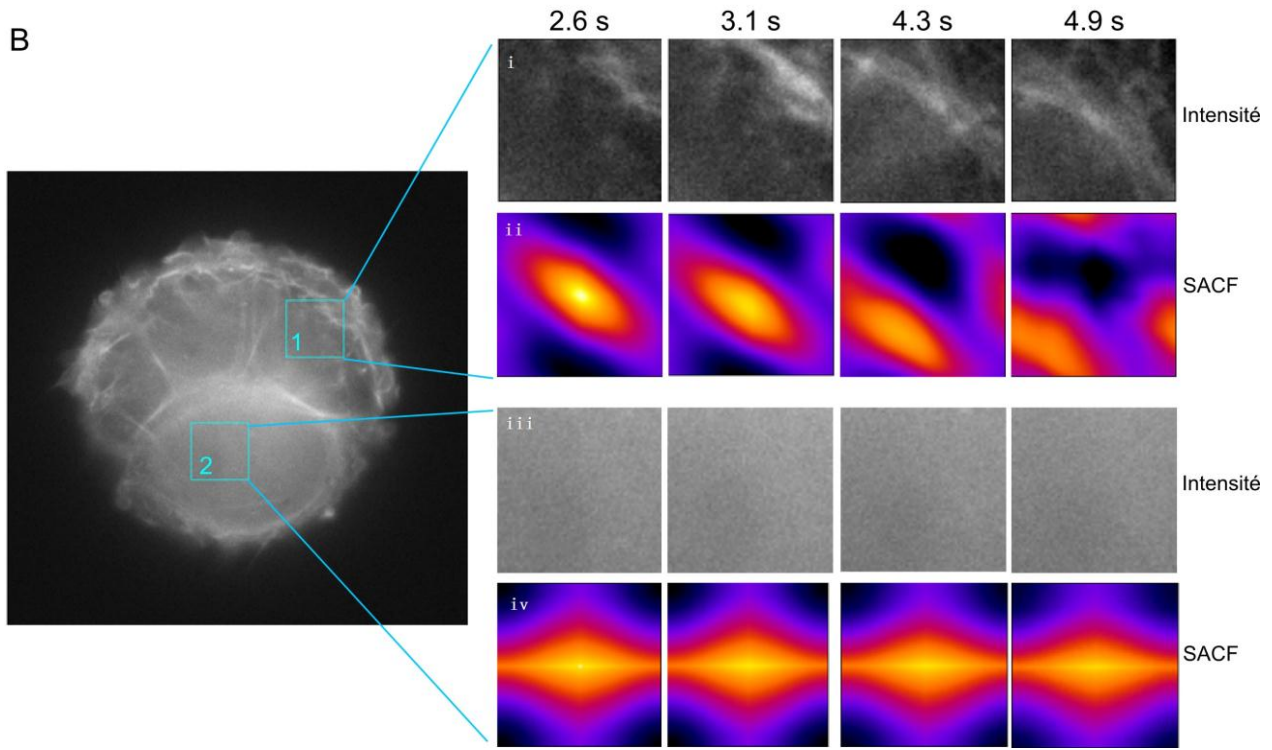

*Figure 4 : Mesure du décalage du pic par des profils en x et y.* (A) *Mesure du décalage du pic par les profils en x et y, la première colonne correspond aux profils de B(ii), la deuxième colonne au profil de B(iii). Les profils sont coupés au milieu de chaque sous-image, dont la longueur correspond à la largeur ou la hauteur d'une sous-image. Les cercles rouges indiquent l'amplitude maximum du profil au cours du temps. La courbe avec les points rouges présente le profil sans décalage. (B) i) un exemple de population flux; et iii) de population diffusion; ii), iv) présente la fonction d'autocorrélation spatiale à différents décalages temporels. La taille de chaque région est 64\*64 pixels.*

### *3.3 Fusion des résultats des sous-séquences:*

A partir de comportements identifiés, on peut appliquer la technique la plus adaptée d'analyse de corrélation d'image pour mesurer les paramètres de la dynamique de la sous-séquence (typiquement TICS pour diffusion et STICS pour flux).

Dans le cas du flux et de l'utilisation du STICS, la mesure de profils en x et y permet également de résoudre un autre problème lors de l'ajustement de la gaussienne de l'équation (2) dans le cas de front d'onde, comme illustré dans la Figure 4 (B) ii). Le modèle n'est pas efficace pour une forme de fonction de corrélation irrégulière, due par exemple à la présence d'un flux en front organisé. Nous proposons de tester les profils en x et y pour mesurer le changement de l'amplitude du pic, comme démontré dans la Figure 4 (A).

Par ailleurs, le résultat est souvent influencé par la taille de la sous-séquence. Pour obtenir un résultat plus précis, l'étape suivante consiste à fusionner les sous-séquences successives pour obtenir des sous-séquences de taille plus importante.

La fusion des résultats est également basée sur le principe de vote, chaque fois le calcul fournit en sortie une réponse comme le vote pour l'un des résultats possibles [4]. Le résultat final est obtenu par la majorité des votes. La règle de vote est suivante:

$$E(x) = \begin{cases} D_i & si \ \sum_{j=1}^{S} e_{i,j} = max_{t=1}^{N} \sum_{j=1}^{S} e_{t,j} \\ Dynamique \ inconnue & sinon \end{cases} \quad (3)$$

Où : $E(x)$ est le résultat de la fusion pour la position x (temporelle ou spatiale) définissant la sous-séquence. $D_i$ est la dynamique identifiée (flux ou diffusion),  S est le nombre de points pris en compte dans la sous-séquence, N présente le nombre de dynamiques (dans notre cas N=2). $e$  représente la sortie (1 ou 0) du vote pour ce résultat.

## 4. Résultats:

Dans cette section, nous montrons les résultats de l'évaluation de notre système. Nous utilisons 2 séquences d'image pour tester notre système d'analyse.

La première séquence est simulée et contient 2 populations : une population de molécules uniques est simulée, avec un déplacement normalement bruité autour d'un mouvement brownien de temps de diffusion imposé pendant 30 images successives, puis qui passent toutes à un déplacement de type flux de vitesse et direction imposée. Ces molécules sont ensuite floutées par une PSF simulée et un faible bruit poissonnien est ajouté.

Dans la Figure 5 nous présentons le résultat de l'analyse de cette séquence. La figure 5(A) montrent l'identification automatique des zones de transitions temporelles. Le panneau B montre les valeurs de diffusion calculées sur les séquences identifiées, correctement, à de la diffusion et pour lesquelles les valeurs calculées sont très proches des valeurs simulées. Le panneau C montrent le résultat approximatif obtenus par profil, mais qui permet de différencier la phase de diffusion de celle de flux. Le panneau D montre la différence entre l'estimation de la vitesse sur la séquence de pur flux (simulé à 0.22 µm/s, valeur calculée 0.28 µm/s) contre la valeur mesurée sur la séquence entière (0.073 µm/s).

Pour la deuxième séquence démontrant l'intérêt sur un système complexe biologique, nous avons choisi de nous intéresser au cytosquelette d'actine. Il s'agit d'un réseau de bio-polymères extrêmement dynamique qui permet à la cellule de maintenir sa forme et d'exercer des forces sur son environnement. La dynamique de ce réseau est l'objet de nombreuses recherches. Dans un fibroblaste polarisé, ce réseau s'assemble au front de migration de la cellule et se propage ensuite vers l'intérieur de la cellule sous l'action de différents partenaires (ADF/cofilin, myosin, …) à une vitesse de quelques µm /minute [5]. Au cours de ce déplacement, ce réseau est modifié et change de dynamique. Le flux d'actine par exemple est plus important sous la membrane de la cellule au front de migration qu'un peu plus au centre de la cellule [6].  Cet objet biologique nous paraissait donc particulièrement indiqué pour la démonstration de l'intérêt de notre méthode qui vise à mettre en évidence des changements spatio-temporels de la dynamique d'une molécule.

Pour visualiser l'actine, nous avons transfecté des cellules fibroblastiques (RPE1) avec le marqueur lifeact [7] couplé à une protéine fluorescente (GFP). Ce système est préférable à de l'actine marquée GFP car cette dernière modifie la dynamique du réseau au sein de la cellule. Les cellules ont ensuite été déposées sur des motifs adhésifs. Ces motifs résultent d'un traitement de surface qui permet de contraindre la forme de la cellule sur une surface de géométrie contrôlée [8]. La cellule ne peut ainsi pas se déplacer. Les flux mesurés sont donc indépendants d'un déplacement global de la cellule et peuvent être interprétés sans que des corrections à ce sujet soient nécessaires.

La figure 6 montre que les vitesses mesurées sont cohérentes avec celles de la littérature (0.01 à 0.06 µm /seconde soit 0.6 à 3 µm /minute), et nous retrouvons visuellement un gradient décroissant de vitesse du front de migration au centre de la cellule. Par ailleurs nous pouvons également considérer l'évolution de la dynamique de certaines zones, alternant flux et diffusion.

Dans les deux séquences, simulée et expérimentale, les profils ont été tracés pour des sous-séquences de 12 images consécutives, de façon sous-échantillonnées pour gagner en temps de calcul : seuls les ACF aux temps relatifs 0, 6 et 11 ont été calculés.

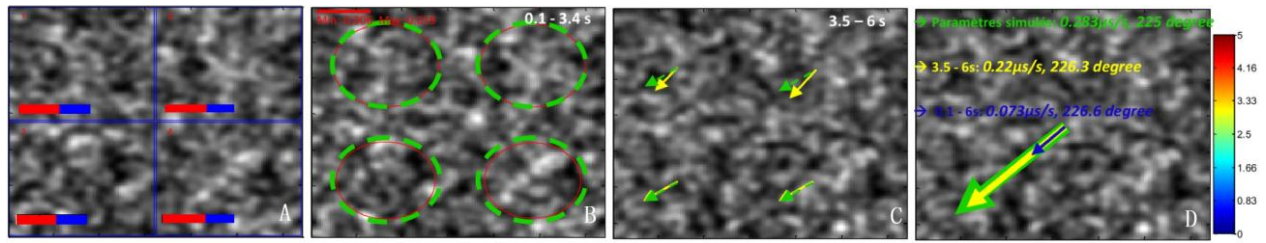

*Figure 5 : Image simulée contenant 2 comportements différents (30 premières images de diffusion et 30 images de flux). (A) 4 sous-régions dans cette séquence de même taille ont été étudiées, l'échelle rouge et bleue montre la proportion des dynamiques identifiées au cours du temps, pour la dynamique et le flux respectivement. (B) et (C) présentent le résultat d'identification, respectivement pour la diffusion et le flux estimé par étude des profils : la transition temporelle a été détectée automatiquement.(D)Résultat final après analyse par STICS sur les régions fusionnées : comparaison entre la valeur simulée (vert) les valeurs trouvées en analysant les régions uniformes automatiquement trouvées (en jaunes) et les valeurs en STICS si elles avaient été calculées sur l'ensemble de la séquence (en bleu).*

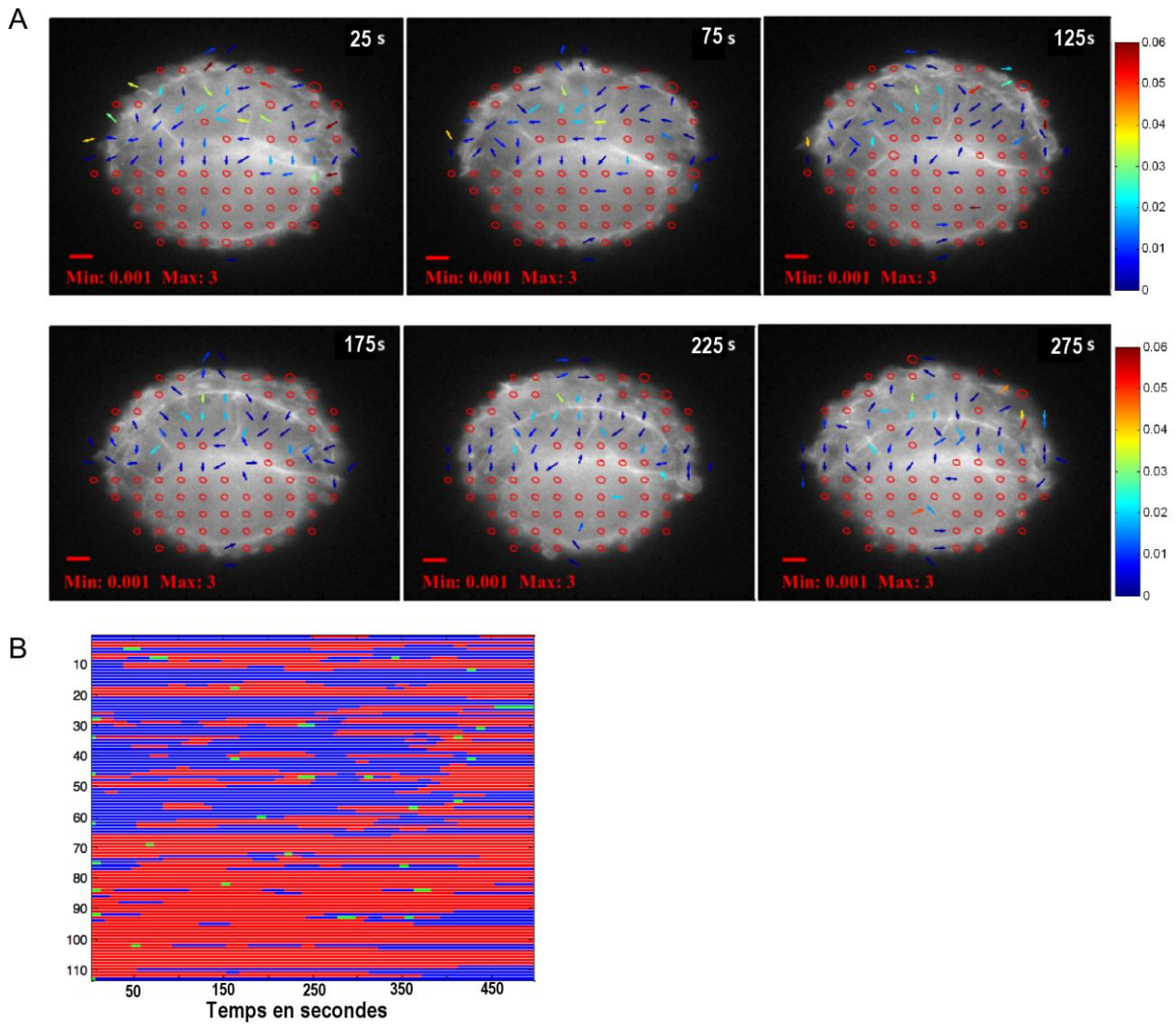

*Figure 6 : (A) Résultat d'identification de la dynamique au cours du temps. Les flèches et les cercles présentent respectivement la dynamique de flux et de diffusion. La barre d'échelle de couleur verticale indique la vitesse de flux codée par couleurs, la barre horizontale présente le coefficient de diffusion codé par le diamètre du cercle. (B) Représentation des différentes dynamiques identifiées consécutivement au cours du temps, chaque ligne correspond à une ROI étudiée, dont la numérotation est donnée sur la figure 3 (F). Le bleu correspond à une dynamique de flux, le rouge à une dynamique de diffusion. Lorsque le vote n'a pas atteint la pluralité, un état 'dynamique inconnue' est attribué, qui est codé par du vert.*

Après identification des dynamiques et estimation des paramètres de mobilité moléculaire, les régions présentant des valeurs similaires peuvent être fusionnées pour définir une région plus grande à comportement uniforme. Un exemple est présenté sur la figure 7.

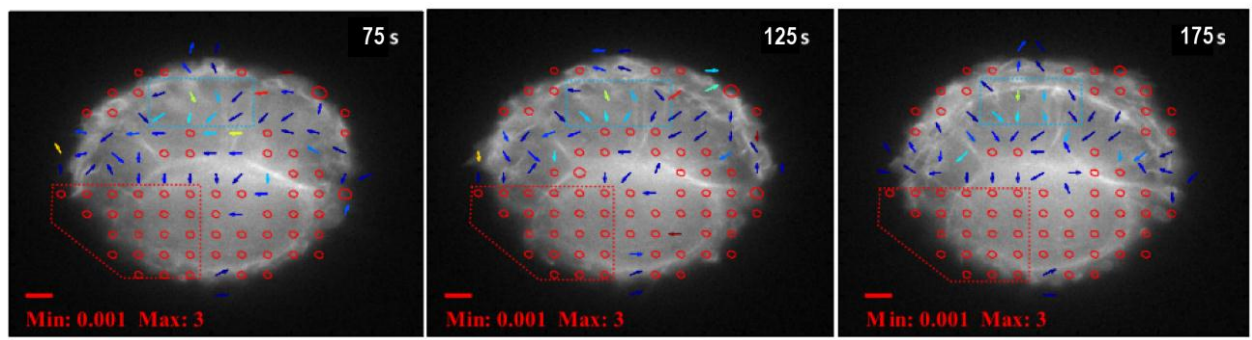

*Figure 7 : Régions identifiées par vote comme présentant le même type de dynamique au cours du temps.*

## 5. Conclusion et perspectives :

Cet article présente un système original d'identification et caractérisation automatique de dynamiques moléculaires différentes au cours du temps par l'utilisation de méthodes type ICS. Cette méthode permet ainsi de définir de façon automatique des régions uniformes spatialement et temporellement pour l'analyse de la dynamique moléculaire. L'utilisation d'images simulées nous a permis de montrer que la différenciation au cours du temps était possible et correcte, et qu'elle permettait de corriger l'estimation des valeurs par rapport aux mesures réalisées sur la séquence entière. En absence de vérité pour la séquence expérimentale, nous pouvons uniquement nous baser sur une vérification visuelle et une comparaison aux données de la littérature. Les perspectives concernent 1) l'amélioration de la précision d'identification dans le changement de dynamique 2) la décomposition des échelles de dynamique dans l'effet de bord, 3) et la fusion des régions à comportement uniforme (même type de mobilité, une valeur du paramètre estimé assez proche). L'étude de modèle mixte (présence de deux populations de dynamique différente dans la même région au même moment) est également une perspective qui apparait essentielle dans ce type de modèle biologique complexe.

## Remerciements

## Bibliographie: